# Synthetic Doubt and the Integrity of Digital Proof before the International Criminal Court

**Law and Political Science Department, University of Tehran, Tehran, Iran**

**Corresponding author: Seyed Mohammad Sadegh Dashti**

**Email: mohammad.dashti@ut.ac.ir**

**ORCID: https://orcid.org/0000-0002-5964-7453**

## Abstract

Generative artificial intelligence complicates evidentiary assessment at the International Criminal Court by enabling fabricated audiovisual material and making authentic digital evidence easier to dismiss as fake. This article calls the latter problem synthetic doubt: an objection alleging that a digital item may be AI-generated or AI-manipulated. Existing scholarship addresses digital evidence, open-source investigations and deepfakes in the ICC context. Yet the question of when synthetic doubt should trigger enhanced reliability review under the Rome Statute, without displacing the prosecution's burden under article 66 or compromising article 67 fair-trial rights, remains underdeveloped. The article proposes a three-stage framework: a baseline authenticity and provenance showing by the proponent; a case-specific articulation threshold for the objector; and proportionate reliability review by the Chamber under articles 64 and 69(4). The framework treats AI-related provenance information and synthetic-media detector outputs as contextual rather than dispositive.

## 1. Introduction

Digital evidence has become central to the proof of international crimes. A smartphone video may capture the aftermath of an airstrike. A social-media post may place an armed unit at a crime scene. Satellite imagery may corroborate the destruction of civilian infrastructure. Metadata may assist with time, location, sequence and source. These materials are attractive because they may survive where witnesses are killed, displaced, intimidated or unavailable; they may also provide a form of contemporaneous documentation that is difficult for perpetrators to suppress entirely.[1] Yet the same characteristics that make digital evidence powerful also make it fragile. Digital traces travel through devices, platforms, compression systems, editing software, reposting practices, archiving tools, cloud services and politically charged information environments before they reach a court.[2]

---

[1] Lindsay Freeman and Rémi Vázquez Llorente, 'Finding the Signal in the Noise: International Criminal Evidence and Procedure in the Digital Age', 19 *Journal of International Criminal Justice* (2021) 163–188; Rafael Braga da

Generative AI adds a distinctive complication. International criminal courts must guard against fabricated or manipulated digital evidence. That concern is serious but familiar in structure: courts have always confronted forged documents, altered photographs, unreliable recordings and contaminated exhibits. The more difficult procedural problem is the reverse: authentic evidence may be dismissed as synthetic simply because realistic fabrication is now widely known to be possible. A video that once appeared self-authenticating may now be accompanied by a familiar allegation: perhaps the face was swapped, the voice was cloned, the file was generated, the metadata was altered or the event was staged through a synthetic composite.

The ICC has already encountered the evidentiary setting in which this problem will arise. The Al-Werfalli arrest warrant in the Libya situation showed the importance of social-media videos to ICC practice and made the authentication of open-source audiovisual material a concrete issue rather than a hypothetical one.[3] Koenig identified the deepfake risk in precisely that context, explaining that such technology could be weaponised to undermine trust in authentic open-source evidence.[4] Vázquez Llorente's practitioner-facing account of deepfakes in the dock confirms that international justice actors now treat generative AI as an operational evidence problem, not merely a speculative technological concern.[5] Freeman and Vázquez Llorente mapped the broader problem of digital evidence in ICC procedure: volume, platform dependence, verification, preservation and judicial evaluation.[6] Braga da Silva, writing in the *International Criminal Law Review*, specifically addressed authentication of digital evidence at the ICC and argued that the Court's existing approach does not fully capture the distinctive authenticity problems of digital material.[7] Hellwig, also in the *International Criminal Law Review*, analysed the potential and challenges of digital evidence in international criminal

---

Silva, 'Updating the Authentication of Digital Evidence in the International Criminal Court', 22 *International Criminal Law Review* (2022) 941–964; Kristina Hellwig, 'The Potential and the Challenges of Digital Evidence in International Criminal Proceedings', 22 *International Criminal Law Review* (2022) 965–988.

[2] OHCHR and Human Rights Center, UC Berkeley, *Berkeley Protocol on Digital Open Source Investigations* (United Nations, New York and Geneva, 2022), pp. 8–12; Sam Dubberley, Alexa Koenig and Daragh Murray (eds.), *Digital Witness: Using Open Source Information for Human Rights Investigation, Documentation, and Accountability* (OUP, Oxford, 2020).

[3] ICC, *Prosecutor v. Mahmoud Mustafa Busayf Al-Werfalli*, ICC-01/11-01/17-2, Pre-Trial Chamber I, Warrant of Arrest (15 August 2017); *see also* Emma Irving, 'And So It Begins: Social Media Evidence in an ICC Arrest Warrant', *Opinio Juris* (17 August 2017), <https://opiniojuris.org/2017/08/17/and-so-it-begins-social-media-evidence-in-an-icc-arrest-warrant/>, accessed 12 August 2026.

[4] Alexa Koenig, '"Half the Truth Is Often a Great Lie": Deep Fakes, Open Source Information, and International Criminal Law', 113 *American Journal of International Law Unbound* (2019) 250–255, at 252–255.

[5] Raquel Vázquez Llorente, 'Deepfakes in the Dock: Preparing International Justice for Generative AI', 20(2) *The SciTech Lawyer* (2024) 28–33.

[6] Freeman and Vázquez Llorente, supra note 1, pp. 163–188.

[7] Braga da Silva, supra note 1, pp. 941–964.

proceedings, including integrity, metadata and technical expertise.[8] More recent empirical work by McDermott, Hausknecht and Liefgreen confirms that ICC judges themselves are attentive to user-generated evidence, AI-edited or AI-generated content and the practical responsibilities of parties when submitting such material.[9]

This article does not claim to be the first to identify deepfakes as a threat to international criminal justice. Nor does it claim to be the first to argue that digital evidence requires careful authentication. Its contribution is narrower. It asks how a Chamber should handle a synthetic-media objection in a way that is procedurally disciplined and compatible with the Rome Statute. The core question is this: when a party says that a particular digital item may be AI-generated or manipulated, what must each actor do? What must the proponent show? What must the objector articulate? What should the Chamber decide? And how can the Court require a case-specific objection without imposing an impermissible reversal of the burden of proof or onus of rebuttal on the accused? The analysis also specifies how AI-related provenance information and synthetic-media detector outputs enter—but do not determine—each stage.

The last question is decisive. Article 67(1)(i) provides that the accused shall not have imposed on him or her any reversal of the burden of proof or any onus of rebuttal.[10] Article 66 places the burden of proving guilt on the Prosecutor and requires conviction beyond reasonable doubt.[11] Any framework that asks an accused to say more than "this could be fake" must therefore be carefully defended. The article's central claim is that a *case-specific articulation threshold* is compatible with those guarantees. Such a threshold does not require the accused to prove falsity, to disprove authenticity or to rebut an element of the charges. It requires only that a party requesting enhanced reliability review identify a concrete reason why the challenged item warrants that review. The threshold is a procedural trigger for judicial case management, not a substantive burden of exculpation.

The article proceeds as follows. Section 2 defines synthetic doubt and distinguishes it from deepfake evidence, ordinary authenticity disputes and reasonable doubt. Section 3 explains why digital evidence before the ICC is vulnerable to both manipulation and denial. Section 4 identifies the relevant Rome Statute architecture: article 69(4), article 64, article 67, article 66, rules 63, 64 and 79, and disclosure-related safeguards. Section 5 reviews the state of the art and specifies the originality claim. Section 6 proposes a three-stage framework. Section 7 applies the framework to ICC and related international criminal practice using analogical scenarios rather than pretending that ICC Chambers have already decided a deepfake objection. Section 8 addresses limitations and institutional implications. Section 9 concludes.

---

[8] Hellwig, supra note 1, pp. 965–988.

[9] Yvonne McDermott, Anne Hausknecht and Alice Liefgreen, 'New Evidence, New Challenges: ICC Judges' Perspectives on User-Generated Evidence and Judging in an Age of Artificial Intelligence', 16(4) *Oñati Socio-Legal Series* (2026) 1571–1595, at 1584–1587, doi:10.35295/osls.iisl.2375.

[10] Rome Statute of the International Criminal Court (adopted 17 July 1998, entered into force 1 July 2002), 2187 UNTS 90, Art. 67(1)(i).

[11] Rome Statute, Arts. 66(2)–(3).

## 2. From Deepfakes to Synthetic Doubt

### 2.1. The Liar's Dividend in Evidentiary Form

Deepfakes are usually discussed as a problem of deception: false media may be admitted or believed as true. That is only half the problem. The deeper procedural concern is that true media may be rejected or discounted as false. Chesney and Citron described this second-order effect as the "liar's dividend": once convincing deepfakes are known to exist, those confronted with authentic evidence may profit from the possibility of fabrication by denying reality.[12] Fallis similarly emphasises that deepfakes threaten knowledge systems not only by adding falsehoods, but by weakening ordinary trust in audiovisual evidence.[13]

In international criminal proceedings, this dividend has a specific evidentiary form. A commander confronted with a video of an unlawful order may say that the audio was cloned. A state confronted with footage of detainee mistreatment may say that the images are fabricated. A militia leader confronted with geolocated social-media material may say that the file was planted or digitally altered. The claim need not ultimately persuade the Chamber to have procedural value. It may delay proceedings, force costly expert disputes, dilute probative value, or provide actors who benefit from uncertainty with a public-facing vocabulary of denial.

This article uses the term *synthetic doubt* to describe that procedural phenomenon. Synthetic doubt is a challenge to digital evidence based on the possibility that the item may be AI-generated, deepfaked, cloned, digitally altered or otherwise synthetic. It is not simply the existence of a deepfake. It is the invocation of the possibility of synthetic manipulation as a reason for the Chamber to discount, exclude, test, delay or re-evaluate a digital item.

### 2.2. Why the Procedural Category Matters

The phrase "deepfake evidence" directs attention to the object: a false or altered exhibit. Synthetic doubt directs attention to the process: an objection or argument that changes the work the court must perform. The distinction matters because courts rarely know at the outset whether the challenged item is fake. A Chamber receiving an objection must decide what level of scrutiny is justified before the truth of the objection is resolved. It must also decide whether the objection affects admissibility, weight, disclosure, expert evidence or only the parties' submissions.

Synthetic doubt is also narrower than ordinary authenticity disputes. Courts have long dealt with forged documents, edited photographs, incomplete recordings and unreliable chain-of-custody evidence. Existing tools remain useful. Pfefferkorn is right to warn against technological panic and to emphasise that legal systems already possess authentication requirements, expert evidence, cross-examination and sanctions.[14] But synthetic doubt places new pressure on those tools. Because the possibility of a deepfake can be invoked in relation to almost any image, video, audio recording or digital communication, courts need a threshold that separates case-specific concerns from generalised technological anxiety.

[12] Bobby Chesney and Danielle Citron, 'Deep Fakes: A Looming Challenge for Privacy, Democracy, and National Security', 107 *California Law Review* (2019) 1753–1820, at 1796–1799.
[13] Don Fallis, 'The Epistemic Threat of Deepfakes', 34 *Philosophy & Technology* (2021) 623–643, at 626–631.
[14] Riana Pfefferkorn, '"Deepfakes" in the Courtroom', 29 *Boston University Public Interest Law Journal* (2020) 245–276, at 268–276.

Nor is synthetic doubt the same as reasonable doubt. Reasonable doubt concerns the final standard for criminal conviction. Synthetic doubt concerns the reliability of a specific item or class of evidence within the record. A Chamber may find that a video is insufficiently reliable without acquitting; it may also admit a video but assign it limited weight. Keeping those inquiries separate is essential. The framework proposed here does not lower the prosecution's burden or require the accused to disprove evidence. It structures the reliability inquiry that precedes, and informs, the Chamber's eventual evaluation of the totality of the evidence.

### 2.3. The Cost Asymmetry

Synthetic-doubt objections have an asymmetric cost structure. Forensic authentication can be expensive, slow and technically demanding. It may require access to original files, platform records, devices, metadata, chain-of-custody documentation, specialist software, laboratory analysis and expert testimony. Well-resourced institutional parties may be able to absorb those costs; defence teams, victims' representatives, local journalists, witnesses and civil society actors may not. De Arcos Tejerizo's analysis of digital evidence and fair-trial rights at the ICC is therefore central: reliability safeguards must not deepen inequality of arms.[15]

The cost asymmetry cuts in both directions. If every digital exhibit must be fully forensically authenticated merely because AI exists, the result will be delay, expense and exclusion of evidence that may be crucial to accountability. If, however, courts refuse to take synthetic-media objections seriously until falsity is proved, parties without access to experts or original files may be unable to challenge genuinely questionable evidence. The appropriate response lies between these extremes: the objection must be specific enough to warrant closer scrutiny, but the threshold must remain low enough that an under-resourced party can trigger review where real indicia of doubt exist.

## 3. Digital Evidence before the ICC

### 3.1. Digital Proof as Ordinary ICC Practice

Digital evidence is no longer peripheral to international criminal justice. It is part of the ordinary evidentiary environment. The Berkeley Protocol provides detailed guidance on identifying, collecting, preserving, analysing and presenting digital open-source information for human rights and international criminal investigations.[16] The Leiden Guidelines translate digitally derived evidence into the procedural context of international criminal courts and tribunals, with particular attention to ICC practice.[17] The Nuremberg Academy's E-Procedure

---

[15] Maria de Arcos Tejerizo, 'Digital Evidence and Fair Trial Rights at the International Criminal Court', 36 *Leiden Journal of International Law* (2023) 749–771, at 762–765.
[16] *Berkeley Protocol*, supra note 2, pp. 2, 17–22, 36–55.
[17] Leiden DDE Project, *Leiden Guidelines on the Use of Digitally Derived Evidence in International Criminal Courts and Tribunals* (2022), pp. 6–12.

Project similarly addresses the collection, preservation, verification, admissibility, weight and fair-trial implications of digital evidence in international criminal proceedings.[18]

These instruments are not binding law. Their importance is practical and evidentiary. They supply a vocabulary for provenance, chain of custody, metadata, hashing, source evaluation, archiving, geolocation, chronolocation, platform effects and presentation. They also show why authenticity cannot be reduced to whether a file looks real. A video may be persuasive because it is embedded in a network of corroboration: original source, collection history, geolocation, time analysis, weather data, satellite imagery, witness testimony, platform history and preservation record. Conversely, a visually convincing item may be weak where source history, metadata and corroboration are absent.

The ICC's practice reflects this broader digital turn. Social-media and video material were central to public discussion of the Al-Werfalli warrant.[19] In *Lubanga*, the Court evaluated video and visual material in connection with the age and presence of alleged child soldiers, illustrating the probative appeal and risk of visual assessment.[20] In other ICC proceedings, telecommunications records, electronic materials, photographs, videos, maps, satellite or aerial imagery and database-derived material have appeared as evidentiary inputs.[21] Adjacent international criminal practice, particularly before the Special Tribunal for Lebanon, shows even more developed use of call-data records and technical telecommunications evidence.[22]

### 3.2. Provenance and Platform Fragility

The integrity of digital evidence depends heavily on provenance. Provenance should be understood broadly: not simply metadata embedded in a file, but the documented history of how the item moved from event to courtroom. That history may include who captured it, where it first appeared, how it was downloaded, whether the original file exists, whether a hash value was generated, what edits or compression occurred, who preserved it, how it was stored, who accessed it, and how the item was presented to the Court.

Platform governance complicates that history. Content moderation may remove atrocity evidence before investigators preserve it. Reposting may detach a file from its original context. Messaging applications may strip metadata. Compression may erase information useful for forensic analysis. Cloud storage and platform APIs may change what is available to investigators. The Berkeley Protocol and Digital Witness literature both emphasise that digital material often reaches investigators after a series of transformations.[23] Synthetic doubt exploits those transformations: missing metadata, lost originals and reposting chains may be innocent, but they can also become reasons for alleging manipulation.

---

[18] International Nuremberg Principles Academy, *E-Procedure: Evidence in Time of Increased Use of Technology and Digitalisation* (Final Report, 2024), pp. 7–9, 28–40.
[19] ICC, *Al-Werfalli*, supra note 3.
[20] ICC, *Prosecutor v. Thomas Lubanga Dyilo*, ICC-01/04-01/06-2842, Trial Chamber I, Judgment (14 March 2012), paras. 131–137; ICC, *Prosecutor v. Thomas Lubanga Dyilo*, ICC-01/04-01/06-3121-AnxA, Appeals Chamber, Judgment on the Appeal against Conviction (1 December 2014), para. 196.
[21] *See generally* Freeman and Vázquez Llorente, supra note 1; Braga da Silva, supra note 1; Hellwig, supra note 1; *Leiden Guidelines*, supra note 17.
[22] Nuremberg Academy, *E-Procedure*, supra note 18, pp. 31–35, discussing call-data records and indicia of authenticity in international criminal practice.

A court that demands perfect provenance risks excluding too much genuine conflict-zone evidence. A court that ignores provenance risks admitting too much evidence on the basis of apparent realism. The legal problem is therefore not to choose between trust and suspicion. It is to decide when provenance weakness, materiality and specific indicia of manipulation justify additional scrutiny.

### 3.3. Detection and Provenance Technologies as Aids, Not Answers

Technical tools can assist, but they cannot decide legal authenticity. NIST's work on synthetic-content risk reduction treats watermarking, provenance, labelling and detection as complementary techniques, not complete solutions.[24] The Coalition for Content Provenance and Authenticity (C2PA) offers a technical standard for documenting content origin and edit history, but a credential cannot itself prove that the underlying event occurred as represented, that the capturing device was trustworthy, or that the item is contextually complete.[25] Conversely, the absence of a C2PA-style credential cannot be fatal for conflict-zone evidence collected by witnesses, victims or journalists using ordinary devices.

AI detection tools are similarly limited. They are probabilistic, often benchmark-sensitive, vulnerable to new manipulation techniques, and capable of false positives where authentic material has been compressed, resized, translated, transcoded or recorded under difficult conditions. Harris's work on synthetic media detection and the burden of proof is valuable precisely because it shows that neither reactive detection nor proactive provenance eliminates the legal question of who must do what when uncertainty remains.[26] A detector output may support a synthetic-doubt objection or respond to one; it should not replace a Chamber's evaluation of materiality, provenance, corroboration, fairness and prejudice.

Generative AI exposes a gap in conventional authentication. A hash and documented chain of custody can show that an acquired file has not changed after collection; they cannot establish that the file was authentic when first created or uploaded. A wholly synthetic file, or one in which only a face, voice, object or short segment has been generated, may still possess a coherent upload and preservation history. At the same time, compression, transcoding and platform transmission may remove metadata or forensic traces, while ordinary smartphone processing may use AI to enhance an image without creating a deceptive deepfake. Conventional provenance therefore remains necessary but is no longer sufficient as the only investigative lens.

---

[23] *Berkeley Protocol*, supra note 2, pp. 44–55; Dubberley, Koenig and Murray, supra note 2.

[24] National Institute of Standards and Technology, *Reducing Risks Posed by Synthetic Content: An Overview of Technical Approaches to Digital Content Transparency*, NIST AI 100-4 (2024), pp. 6–7, 38–39.

[25] Coalition for Content Provenance and Authenticity, *C2PA Technical Specification*, version 2.4 (April 2026), <https://spec.c2pa.org/specifications/specifications/2.4/specs/C2PA_Specification.html>, accessed 12 August 2026.

[26] K. R. Harris, 'Synthetic Media Detection, the Wheel, and the Burden of Proof', 37 *Philosophy & Technology* (2024), Art. 131; K. R. Harris, 'AI or Your Lying Eyes: Some Shortcomings of Artificially Intelligent Deepfake Detectors', 37 *Philosophy & Technology* (2024), Art. 7.

Technical indicators relevant to alleged AI generation or modification fall into four related but distinct categories: signed provenance metadata, including C2PA manifests; provider watermarks or recoverable signals; platform or provider labels; and passive, content-based detector outputs. A manifest should be examined in stages: whether it is well-formed, valid and, where relevant, trusted; what assertions it makes about creation, editing and digital source type; and whether those assertions are legally material and substantively credible. C2PA 2.4 includes source-type values such as “trainedAlgorithmicMedia” and “compositeWithTrainedAlgorithmicMedia”, but successful validation can show that assertions are cryptographically bound and have not been altered; it does not establish that the signer, device or assertion is trustworthy, or that the depicted event occurred as represented. Watermarks and labels likewise require source-specific interpretation: a label may reflect self-disclosure, signed provenance or a proprietary classifier. Their absence, especially from older, reposted or conflict-zone material, is ordinarily neutral rather than evidence of fabrication.

Passive detectors have narrower use cases. They may support investigative triage, help formulate a Stage Two objection, identify a region, frame or voice segment for focused examination, or inform a Stage Three expert review. Their output should not be treated as a calibrated probability of falsity. Reliability depends on the examined modality, the exact input and any preprocessing, the model and version, the decision threshold, performance on materially comparable data, false-positive and false-negative rates, exposure to novel generators, and robustness to compression, resizing, noise and other benign or adversarial changes. The value of an alert also depends on reproducibility and, where proportionate, independent confirmation. A positive result may justify scrutiny but does not prove fabrication; a negative result does not prove authenticity. These are non-exhaustive factors under article 69(4), not a freestanding technical admissibility test.

## 4. The Rome Statute Architecture

### 4.1. Article 69(4) as the Primary Reliability Anchor

Article 69(4) is the central provision for synthetic-doubt disputes. It authorises the Court to rule on the relevance or admissibility of any evidence, taking into account, among other things, probative value and any prejudice the evidence may cause to a fair trial or to a fair evaluation of witness testimony.[27] ICC Chambers have treated admissibility through a flexible inquiry into relevance, probative value and prejudicial effect, with reliability informing probative value.[28]

[27] Rome Statute, Art. 69(4).

[28] ICC, *Prosecutor v. Thomas Lubanga Dyilo*, ICC-01/04-01/06-1399, Trial Chamber I, Decision on the Admissibility of Four Documents (16 June 2008), para. 26; ICC, *Prosecutor v. Jean-Pierre Bemba Gombo*, ICC-01/05-01/08-1386, Trial Chamber III, Decision on the Admission of Materials into Evidence (3 May 2011), para. 10.

That structure is well suited to digital evidence because it avoids a rigid categorical rule while still requiring reasoned evaluation.

Article 69(7) should not be stretched into the primary rule for synthetic doubt. It excludes evidence obtained by means of a violation of the Rome Statute or internationally recognised human rights where the violation casts substantial doubt on reliability or the admission of the evidence would seriously damage the integrity of proceedings.[29] Its trigger is a rights violation in the manner of obtaining evidence. A digital item may be unreliable because of manipulation, weak provenance or missing metadata without having been obtained through a rights violation. In such cases, article 69(4), not article 69(7), is the correct starting point.[30]

Article 69(4) is not, however, the whole answer. Synthetic doubt also engages the case-management powers in article 64, the defence rights in article 67, the presumption of innocence under article 66, disclosure obligations and the Rules of Procedure and Evidence. The correct doctrinal position is therefore not that article 69(4) is sufficient in isolation, but that it is the primary admissibility and reliability anchor within a broader procedural architecture.

### 4.2. Article 64 and Case Management

Article 64(2) requires the Trial Chamber to ensure that the trial is fair and expeditious and conducted with full respect for the rights of the accused and due regard for the protection of victims and witnesses. Article 64(8)(b) empowers the Chamber to rule on the admissibility or relevance of evidence.[31] Rules 63 and 64 reinforce the Court's flexible approach to evidence and provide the procedure for objections concerning relevance or admissibility.[32]

These provisions supply the basis for requiring a party to explain why a synthetic-media objection warrants enhanced review. Courts manage proceedings by requiring objections to be articulated. A party objecting to relevance, prejudice, privilege, disclosure, authenticity or expert methodology is ordinarily expected to state the basis of the objection. The articulation requirement proposed here is an application of that ordinary case-management logic to the specific problem of AI-era digital authenticity.

### 4.3. Article 67(1)(i), Article 66 and the No-Onus Constraint

The most sensitive issue is article 67(1)(i). It provides that the accused is entitled not to have any reversal of the burden of proof or any onus of rebuttal imposed.[33] Article 66 confirms that the accused is presumed innocent, that the Prosecutor bears the burden of proof, and that conviction requires proof beyond reasonable doubt.[34] A framework for synthetic doubt must therefore avoid turning the accused's objection into a duty to prove falsity.

---

[29] Rome Statute, Art. 69(7).

[30] ICC, *Prosecutor v. Thomas Lubanga Dyilo*, ICC-01/04-01/06-1981, Appeals Chamber, Judgment on the Appeals Relating to Evidence Seized in Beni (11 September 2008), paras. 52–56.

[31] Rome Statute, Arts. 64(2), 64(8)(b).

[32] ICC Rules of Procedure and Evidence, ICC-ASP/1/3 and subsequent amendments, Rules 63–64.

[33] Rome Statute, Art. 67(1)(i).

[34] Rome Statute, Arts. 66(1)–(3).

The case-specific articulation threshold avoids that result for four reasons. First, it does not determine guilt, responsibility, an element of the crime or an affirmative defence. It determines only whether a particular evidentiary objection is sufficiently specific to trigger enhanced reliability review. Second, it requires articulation rather than proof. The objector need not show that the item is fake; the objector need only identify a concrete feature of the item, its provenance or its context that justifies scrutiny. Third, failure to meet the threshold does not conclusively validate the evidence. The item remains subject to ordinary weight assessment, cross-examination, disclosure arguments and the prosecution's ultimate burden. Fourth, the threshold is symmetrical: when the prosecution challenges defence evidence on synthetic-media grounds, it must also articulate concrete reasons.

Rule 79 provides an important analogy. It requires the defence to notify the Prosecutor and the Chamber if it intends to raise an alibi or a ground for excluding criminal responsibility.[35] That notice obligation does not transfer to the accused the burden of proving innocence. It allows the Court and the opposing party to manage proceedings fairly and effectively. The analogy is limited: synthetic doubt concerns evidence rather than defences. But it demonstrates the broader point that not every procedural obligation placed on the accused is a prohibited onus of rebuttal. A procedural notice or articulation requirement becomes constitutionally problematic only if it effectively requires the accused to establish a substantive fact as the price of avoiding conviction. The framework proposed here does not do that.

### 4.4. Disclosure, Equality of Arms and Expert Evidence

Synthetic-doubt disputes cannot be separated from disclosure and equality of arms. If the prosecution relies on a video, audio recording or digital communication, the defence may need access to the original file, metadata, collection notes, preservation records, expert methodology and chain-of-custody information in order to articulate a meaningful challenge. Article 67(1)(e) protects the accused's ability to examine witnesses and obtain the attendance and examination of witnesses under the same conditions as the prosecution.[36] Article 67(2) requires disclosure of exculpatory or material evidence. A Chamber should not reject a synthetic-doubt objection as insufficient where the insufficiency results from the opposing party's control of the relevant source material.

Expert evidence presents a related problem. Gillett and Fan emphasise that digital open-source information may require expert presentation while maintaining rigorous fact-finding standards.[37] Where a Chamber orders or admits expert analysis of synthetic-media risk, parties must be able to test the expert's methodology, data, assumptions, error rates, limitations and independence. An expert label cannot turn probabilistic detection into certainty. The legal question remains whether, in context, the item is sufficiently reliable and whether reliance on it would be fair.

---

[35] ICC Rules of Procedure and Evidence, Rule 79.
[36] Rome Statute, Arts. 67(1)(e), 67(2).
[37] Matthew Gillett and Wallace Fan, 'Expert Evidence and Digital Open Source Information: Bringing Online Evidence to the Courtroom', 21 *Journal of International Criminal Justice* (2023) 661–693.

## 5. State of the Art and the Original Contribution

The relevant scholarship now falls into four clusters. The first is ICC digital-evidence scholarship. Freeman and Vázquez Llorente provide the foundational account of the digital-evidence landscape at the ICC.[38] Braga da Silva focuses specifically on authentication and the need to update the ICC's approach to digital evidence.[39] Hellwig analyses digital evidence in international criminal proceedings and the challenges of metadata, expertise and reliability.[40] De Arcos Tejerizo places the problem within fair-trial rights and equality of arms.[41] Aksamitowska demonstrates that digital evidence in core-crimes prosecutions is not confined to The Hague and that domestic cases also face verification and authentication difficulties.[42]

The second cluster concerns open-source investigations and digital integrity. The Berkeley Protocol, Leiden Guidelines, Digital Witness and the Nuremberg Academy E-Procedure materials provide practical and procedural guidance for preservation, verification, chain of custody, presentation, admissibility and weight.[43] These instruments do not dictate the legal test, but they shape what a reasonable baseline showing can include.

The third cluster concerns deepfakes, AI-generated evidence and detection. Koenig first connected deepfakes, open-source information and international criminal law through the Al-Werfalli scenario.[44] Pfefferkorn warns against overreaction and argues that existing tools retain value.[45] Delfino favours more active judicial gatekeeping in response to deepfake risk.[46]

Harris focuses on synthetic-media detection and burden allocation, showing why legal procedure must decide who bears what obligation when technical certainty is unavailable.

Within this cluster, Hausknecht supplies the most direct operational account for user-generated evidence. Her analysis identifies three linked weaknesses in detector outputs: the representativeness and quality of training data; the absence of an agreed classification threshold; and reduced certainty when a detector encounters unfamiliar material. Those weaknesses are acute for evidence of international crimes, which is often low-resolution, noisy, non-English and captured in conflict settings. At the same time, ordinary smartphone AI functions may produce content that is neither conventionally "real" nor a deceptive deepfake. Hausknecht accordingly favours transparent methodology, chain-of-custody documentation, expert scrutiny, corroboration and a combination of human analysis with machine tools. The present article takes the next doctrinal step: it translates those operational limits into differentiated responsibilities for the proponent, objector and Chamber, and treats the cost of rigorous authentication as a potential equality-of-arms concern.[47]

---

[38] Freeman and Vázquez Llorente, supra note 1.
[39] Braga da Silva, supra note 1.
[40] Hellwig, supra note 1.
[41] De Arcos Tejerizo, supra note 15.
[42] Karolina Aksamitowska, 'Digital Evidence in Domestic Core International Crimes Prosecutions', 19 *Journal of International Criminal Justice* (2021) 189–211.
[43] *Berkeley Protocol*, supra note 2; *Leiden Guidelines*, supra note 17; Nuremberg Academy, *E-Procedure*, supra note 18; Dubberley, Koenig and Murray, supra note 2.
[44] Koenig, supra note 4.
[45] Pfefferkorn, supra note 14.
[46] Rebecca A. Delfino, 'Deepfakes on Trial: A Call to Expand the Trial Judge's Gatekeeping Role to Protect Legal Proceedings from Technological Fakery', 74 *Hastings Law Journal* (2023) 293–342, at 311–325.

Building on those insights, this article moves from general burden-allocation theory to a Rome Statute-specific mechanism: an articulation threshold grounded in article 64, tested against article 67(1)(i), and operationalised through article 69(4). NIST and C2PA materials show that technical provenance tools can support, but not replace, legal assessment.[48]

The fourth cluster concerns the liar's dividend and the epistemic environment of denial. Chesney and Citron identify the strategic benefit of denying authentic evidence after deepfakes become plausible.[49] Fallis explains why deepfakes weaken trust in audiovisual knowledge systems.[50] This literature explains why the problem is not just false evidence, but the strategic production of doubt around true evidence.

The article's originality is specific. It does not claim to be the first to notice deepfakes, the first to discuss digital evidence at the ICC, or the first to call for better authentication. It claims that the state of the art lacks a Rome Statute-specific framework for determining when synthetic-media objections should trigger enhanced reliability review without imposing an impermissible onus on the accused. The key doctrinal contribution is the case-specific articulation threshold: a procedural trigger grounded in article 64, channelled through article 69(4), constrained by article 67(1)(i), and sensitive to disclosure and equality of arms.

## 6. A Three-Stage Framework for Synthetic Doubt

The framework has three stages. It is deliberately modest. It does not create a new exclusionary rule. It does not presume that deepfake detection technology can resolve authenticity disputes. It does not require full forensic authentication of every digital item. It does not reverse the prosecution's burden of proof. It structures the Chamber's inquiry so that synthetic doubt is neither ignored nor allowed to become an evidentiary veto. Table 1 summarises the allocation of obligations; the subsections that follow develop each stage.

**Table 1. Three-Stage Framework for Synthetic Doubt**

| Stage | Actor | Core obligation | Legal and methodological anchors | Consequence |
|---|---|---|---|---|

[47] Harris, 'Synthetic Media Detection', supra note 26; Anne Hausknecht, 'The Impact of Deepfakes on Trust in User-Generated Evidence', in Thomas D. C. Bennett and Rebecca Moosavian (eds.), *Deepfakes and the Law: Challenges, Responses, and Critique* (Routledge, Abingdon, forthcoming September 2026), preprint <https://papers.ssrn.com/sol3/papers.cfm?abstract_id=5352911>, accessed 12 August 2026.
[48] NIST AI 100-4, supra note 24; *C2PA Technical Specification*, supra note 25.
[49] Chesney and Citron, supra note 12.
[50] Fallis, supra note 13.

| Stage | Actor | Core obligation | Legal and methodological anchors | Consequence |
|---|---|---|---|---|
| 1 | Proponent | Provide a proportionate baseline showing of authenticity, ordinary provenance and, where reasonably available, AI-related provenance information, labels, watermarks and known AI-assisted transformations. | Article 69(4); Berkeley Protocol; Leiden Guidelines; NIST; C2PA | Item enters ordinary relevance, probative-value and fairness assessment; weak baseline may reduce weight or support exclusion. |
| 2 | Objector | Identify case-specific reasons for synthetic doubt, including concrete provenance inconsistencies or sufficiently documented AI-related technical signals; absence of credentials alone is insufficient. | Article 64 case management; Article 67(1)(i) constraint; NIST; C2PA | If threshold is not met, no enhanced reliability review is triggered; ordinary weight arguments remain available. |
| 3 | Chamber | Conduct proportionate reliability review, evaluating provenance information and detector outputs through context-specific, non-exhaustive factors alongside ordinary corroboration and fairness. | Article 69(4); Article 67(1)(e); Rules 63–64; NIST | Chamber may order disclosure, expert evidence, focused questioning, corroboration, reduced weight or exclusion where necessary. |

### 6.1. Stage One: Baseline Showing by the Proponent

The first stage belongs to the party offering the digital item. The proponent must provide a baseline showing sufficient for the Chamber and opposing parties to understand what the item is, where it came from, how it was collected, how it was preserved, and why it is sufficiently reliable for the purpose for which it is offered. The standard is a proportionate foundation, not technical perfection.

In the generative-AI context, the baseline has two layers. Conventional provenance remains essential. In addition, where reasonably available, the proponent should preserve and disclose any C2PA manifest or external provenance record; any provider watermark or platform or provider label; and any known AI-assisted generation, enhancement or transformation. This is a duty to preserve and explain existing indicators, not to create credentials retrospectively or to screen every exhibit with a detector. The absence of credentials is ordinarily neutral, particularly for older, reposted or conflict-zone evidence; its significance depends on whether such credentials should reasonably have existed and whether there is evidence that they were removed or contradicted.

The required showing will vary with materiality. A peripheral social-media image used only for background should not require the same foundation as a recording said to capture a command to commit crimes. A video that is cumulative of multiple witness accounts should not require the same degree of scrutiny as a lone, decisive audiovisual exhibit. The baseline may include source information, original URL, device or platform data, upload history, preservation records, hash values, metadata, chain-of-custody documentation, geolocation, chronolocation, witness testimony, expert analysis and corroboration.[51]

The baseline requirement applies to all parties, but its application must be resource-sensitive. Prosecutors and institutional investigators ordinarily have stronger preservation capacity than displaced witnesses, civil society actors or defence teams. That does not mean defence evidence should be admitted uncritically. It means that the Chamber should evaluate the adequacy of the baseline showing in light of the party's access to the item, its source and its preservation environment.

### 6.2. Stage Two: Case-Specific Articulation by the Objector

The second stage belongs to the party raising synthetic doubt. The objector must identify concrete, case-specific reasons why the item warrants enhanced reliability review. A bare assertion that "AI exists" or that "videos can be deepfaked" is insufficient. That assertion states a general technological possibility, not a case-specific objection.

AI-related technical indicators may satisfy the articulation threshold when sufficiently concrete: an invalid or internally inconsistent manifest; a source declaration or platform label that conflicts with the account given of the item's origin or editing history; a recoverable watermark associated with synthetic generation; a credible, modality-specific detector alert tied to the examined file or segment; or an unexplained AI-enabled transformation. Neither the absence of a credential nor an unsupported screenshot of a detector score should suffice by itself. At this stage, however, an under-resourced objector need not supply complete validation data; a documented preliminary signal may justify disclosure or focused review, with full methodological testing reserved for Stage Three.

Concrete reasons may include missing original files, unexplained compression or editing, broken chain of custody, inconsistent metadata, multiple inconsistent versions, visible or audible anomalies, absence of corroboration for a central item, credible preliminary expert concern, evidence of coordinated manipulation linked to the source, or contextual information showing that the same source has disseminated manipulated material. The objector need not prove manipulation; it must identify a factual predicate that makes closer scrutiny proportionate.

The threshold should be lower where the objector lacks access to material controlled by the proponent or a platform. If the defence cannot access the original file, metadata or collection notes because those materials remain in the prosecution's possession, the Chamber should consider disclosure or inspection before treating the objection as insufficient. A threshold designed to prevent speculative objections should not become a shield for information asymmetry.

In practice, the articulation threshold would not require a new procedural instrument. It could be applied through written submissions on admissibility or weight, responses to a party's notice of reliance on digital evidence, trial-management conferences or status conferences, focused disclosure applications, or oral objections when the evidence is tendered. The Chamber's authority to organise those steps follows from article 64 and rules 63–64; the threshold is therefore a way of structuring existing procedural discretion, not a free-standing burden invented outside the statutory framework.

[51] *Berkeley Protocol*, supra note 2, pp. 36–55; *Leiden Guidelines*, supra note 17, pp. 6–12.

### 6.3. Stage Three: Proportionate Reliability Review by the Chamber

Once Stage Two is satisfied, the Chamber conducts proportionate reliability review. The review should be calibrated to materiality, seriousness of the doubt, available corroboration, technical complexity, resource burden and fairness. It is not automatically a mini-trial on authenticity.

When provenance information or detector outputs are relied upon, the Chamber should ask proportionate questions directed to fitness for use. For a manifest, it should ask whether the manifest is technically valid and bound to the asset, what it asserts, who signed it, and whether the signer is trusted for the relevant purpose. For a detector, the Chamber should consider the exact file and hash examined; the modality, region or segment analysed; the tool, model, version and any preprocessing; the score and decision threshold; and false-positive and false-negative performance on materially comparable data, including compressed, noisy or unfamiliar-generator content. It should also consider whether the result can be reproduced or independently confirmed. These are non-exhaustive reliability factors, applied where relevant and reasonably available. They do not create an ICC analogue of a rigid technical gatekeeping rule.

At the low end, the Chamber may require the proponent to provide the best available version of the file, metadata, collection notes, chain-of-custody records or expert methodology. At the middle level, it may permit focused questioning, order limited expert evidence, or require corroboration. At the high end, where the item is central and the specific doubt is serious, unresolved and prejudicial, the Chamber may reduce weight substantially or exclude the item under article 69(4). Exclusion should remain exceptional; weight, disclosure and targeted expert review will often be more proportionate.

The Chamber should state expressly what the objection affects: admissibility, weight, disclosure, expert evidence or trial management. That discipline matters because synthetic doubt otherwise risks becoming a confused mixture of technical anxiety and legal conclusion. A Chamber may admit an item but assign limited weight. It may require disclosure without making an admissibility ruling. It may reject a speculative objection while preserving the accused's right to challenge weight at the close of the case. The legal consequence should match the reason for doubt.

### 6.4. Confirmation, Trial, Appeal and Reparations

The framework is designed primarily for trial, where evidence is assessed in relation to guilt or innocence and article 66's beyond-reasonable-doubt standard remains decisive. But the same logic can apply, with appropriate adjustments at other procedural stages. At confirmation, the Chamber is concerned with whether there are substantial grounds to believe that the person committed each charged crime. The articulation threshold should therefore be lighter than at trial: it should screen out bare technological speculation while avoiding an elaborate authenticity inquiry that would distort the limited function of confirmation. At the arrest-warrant or summons stage, as illustrated by Al-Werfalli-type reliance on social-media video, the threshold should be lighter still, but a Chamber should still record whether reliance on digital material rests on source history, corroboration, preservation method or other reliability indicators.

At trial, by contrast, the Chamber can demand a more developed showing because the evidence may contribute to findings of criminal responsibility. On appeal, synthetic doubt will usually arise as a challenge to whether the Trial Chamber reasonably assessed reliability, fairness or prejudice, rather than as a fresh authenticity hearing. At reparations, the framework may assist where digital evidence is used to identify harm, location, destruction or victim participation, but the criminal burden of proof is no longer the same.

The framework is therefore not a single technical standard. It is a procedural discipline that varies by stage: identify the evidentiary role of the item, require a proportionate baseline showing, demand case-specific articulation for enhanced review, and select a response suited to the procedural moment.

## 7. Application to ICC and Related Practice

No ICC Chamber has yet decided a fully developed synthetic-media objection. The following examples are therefore analogical. They do not claim that the cases involved deepfakes. They test how the proposed framework fits the types of digital or digitally derived evidence that international criminal courts have actually encountered.

### 7.1. Al-Werfalli: Social-Media Video and the Risk of Denial

The Al-Werfalli warrant remains the most vivid ICC example because it involved alleged executions documented through video and social-media material in the Libya situation.[52] Commentators quickly recognised the significance of the warrant for open-source evidence and the evidentiary risks that would arise if such material were later challenged at trial.[53] Koenig used the case to illustrate the deepfake problem: if authentic social-media evidence becomes contestable through synthetic-media allegations, international criminal justice must be prepared to explain why some digital material is reliable enough to use.[54]

Under Stage One, the proponent of Al-Werfalli-type evidence would need to show the source, upload history, preservation method, geolocation, chronolocation, chain of custody and corroborating information. Under Stage Two, the accused or another objector could not simply say that execution videos can be fabricated. The objector would need to identify a specific issue: missing originals, inconsistent versions, unexplained editing, visible anomalies, unreliable source history, or expert concern. Under Stage Three, the Chamber could respond by ordering disclosure of collection records, admitting the videos with caution, requiring expert analysis, seeking corroboration, or reducing weight.

[52] ICC, *Al-Werfalli*, supra note 3.

[53] Irving, supra note 3; Alex Whiting, 'The ICC's New Libya Case: Extraterritorial Evidence for an Extraterritorial Court', *Just Security* (23 August 2017), <https://www.justsecurity.org/44383/iccs-libya-case-extraterritorial-evidence-extraterritorial-court/>, accessed 12 August 2026.

[54] Koenig, supra note 4, pp. 252–255.

AI-related technical indicators would refine, not displace, that analysis. A valid C2PA manifest recording capture and subsequent edits could strengthen the Stage One foundation, but would not establish that the executions occurred as represented. Its absence from 2017 social-media footage would be neutral. At Stage Two, by contrast, a manifest validation failure, a label indicating an AI source that conflicts with the account given of the item, a recoverable generative watermark or a documented detector alert directed to a particular face, frame or audio segment could warrant review. At Stage Three, the Chamber would examine the indicator's origin and meaning and, for a detector output, the input, version, threshold, performance on comparable validation data and reproducibility before considering the result alongside source history and corroboration.

This example shows the practical point of the framework. It avoids two errors: treating social-media videos as self-authenticating because they are vivid, and allowing a bare deepfake allegation to neutralise them because generative AI exists.

### 7.2. Lubanga: Visual Evidence and the Limits of Appearance

*Lubanga* illustrates the need to assess visual material without equating visual force with legal certainty. The Trial Chamber assessed evidence that included videos and other visual material in relation to child soldiers and age; the Appeals Chamber later addressed the reasonableness of reliance on visual assessment in circumstances where other evidence was limited.[55] This is not a deepfake case. Its relevance lies in the evidentiary logic: visual material can be probative, but courts must explain how they move from image to finding.

If a synthetic-doubt objection were raised to comparable video evidence today, Stage One would require the proponent to provide a provenance and reliability foundation proportionate to the video's role. Stage Two would ask whether the objector can identify concrete reasons for doubt: missing originals, uncertain source, inconsistent metadata, signs of editing, or lack of corroboration for a central inference. Stage Three would permit a range of responses. If the video is cumulative and corroborated, weight analysis may suffice. If the video is central and weakly sourced, enhanced disclosure or expert assistance may be proportionate. The framework therefore translates *Lubanga*'s lesson into the AI era: appearance can contribute to proof, but it should not substitute for structured reasoning about reliability.

### 7.3. Geospatial Material: Institutional Provenance and Analytical Transparency

Geospatial and satellite material illustrate a different problem. Such evidence often has stronger institutional provenance than user-generated content: it may originate from commercial providers, state sources, humanitarian organisations or specialist analysts. But it still requires transparent methodology. A satellite image may show destruction, movement, terrain or proximity; the legal inference from the image depends on the image's date, resolution, enhancement, comparison, interpretation and expert explanation.[56]

---

[55] *Lubanga* Trial Judgment, supra note 20, paras. 131–137; *Lubanga* Appeals Judgment, supra note 20, para. 196.

Under the framework, the Stage One showing for geospatial material may be stronger where source, date, provider and image-processing history are documented. The Stage Two threshold should therefore require a more specific objection than in the case of an anonymous social-media video. It may be satisfied by evidence of inconsistent imagery from the same period, uncertainty about enhancement, disputed geolocation, manipulation of overlays, or unsupported expert inference. Stage Three would typically involve expert clarification rather than exclusion. The question is not whether satellite imagery can be manipulated in the abstract; it is whether this image, processed and interpreted in this way, supports the inference for which it is offered.

### 7.4. Telecommunications and Digital Records

Telecommunications evidence, call-data records and digital communications present a third category. They may carry institutional indicia of reliability—corporate records, provider watermarks, extraction logs, database architecture, internal consistency and expert testimony—but they can also raise questions of attribution, completeness, chain of custody and interpretation. The Nuremberg Academy's E-Procedure materials note that international criminal practice has developed indicia for assessing the authenticity of call-data records, including provider markings, correspondence with timelines, expert testimony and chain-of-custody information.[57]

Synthetic doubt in this context may not look like a face-swap or voice clone. It may take the form of an allegation that records were altered, misattributed, incompletely extracted, or generated from unreliable databases. Stage One would require the proponent to show how the records were obtained, preserved and interpreted. Stage Two would require a concrete challenge to the extraction, provider source, metadata, attribution or chain of custody. Stage Three could involve disclosure of extraction methodology, expert evidence, cross-checking against other records, or reduced weight where attribution remains uncertain.

### 7.5. Rule 79 and the Procedural-Obligation Analogy

The Rule 79 analogy helps answer the constitutional objection. The ICC already recognises that the defence may be required to give notice of certain procedural positions, such as an alibi or a ground excluding criminal responsibility.[58] Such notice does not reverse the burden of proof; it allows the Chamber and Prosecutor to organise proceedings fairly. The synthetic-doubt articulation threshold performs a comparable function at the evidentiary level. It does not require the accused to prove falsity; it requires the objector to identify the basis for asking the Chamber to devote resources to enhanced review.

---

[56] Hellwig, supra note 1, pp. 965–988; Nuremberg Academy, *E-Procedure*, supra note 18, pp. 28–40; *Leiden Guidelines*, supra note 17.
[57] Nuremberg Academy, *E-Procedure*, supra note 18, pp. 31–35.
[58] ICC Rules of Procedure and Evidence, Rule 79.

The analogy also clarifies the limits of the threshold. If an articulation requirement becomes so demanding that the accused must produce expert evidence before being heard, it would approach the forbidden territory of article 67(1)(i). If, however, the threshold requires only a concrete factual predicate and remains sensitive to disclosure asymmetry, it stays within case-management authority. The Chamber's task is to police that line.

**8. Limits and Institutional Implications**

The proposed framework has limits. It cannot solve staged events, where the recording is authentic but the event itself was orchestrated. It cannot detect all selective editing. It cannot prevent false provenance credentials. It cannot recover evidence deleted before preservation. It cannot make AI detectors reliable across all languages, accents, compression environments and recording conditions. It cannot ensure that all defence teams have equal access to experts. It also does not address the distinct question of AI-assisted investigative analysis, such as algorithmic geolocation, image enhancement, pattern detection or translation used by investigators before evidence is presented.

Those limits are not defects in the framework; they define its function. The framework is not a technology of truth. It is a legal procedure for deciding when doubts about synthetic manipulation are specific enough to justify additional reliability work, and what form that work should take.

Several institutional implications follow. First, investigators should preserve original files and generate hash values as early as possible. Second, parties should document collection methodology in a way that can be disclosed or summarised without endangering witnesses and sources. Third, Chambers should distinguish explicitly between admissibility, weight and disclosure consequences. Fourth, the Court should have access to neutral technical expertise without delegating legal judgment to experts. Fifth, judicial reasoning should avoid both technological complacency and technological panic. The Court's legitimacy in contested information environments depends on explaining why it trusts, distrusts or conditionally relies on digital evidence.

Where AI generation or modification is in issue, preservation requires a more precise audit trail. Investigators should preserve native files and submitted derivatives separately; retain embedded manifests and any externally linked provenance records or sidecars; hash the native asset before analytical processing; document known AI-assisted enhancement, translation, noise reduction or other transformation; and record, for any detector analysis, the exact input, preprocessing, date, settings, tool, model and version, and output. Where platforms supply labels or provenance information, investigators should preserve the label, its source and the method by which it was obtained. Technical protocols should be reviewed periodically as generators, standards and detectors change. These practices increase transparency without making modern credentials a condition of admissibility.

A final implication concerns public communication. International criminal judgments speak beyond the courtroom. They are read by affected communities, states, civil society, victims, perpetrators and audiences already exposed to disinformation. If courts ignore synthetic-media risks, they may appear naive. If they treat all digital evidence as suspect, they may weaken accountability for crimes that are documented primarily through digital traces. A structured articulation-and-review framework helps courts explain the difference between genuine technological concern and strategic denial.

## 9. Conclusion

Synthetic doubt will become a recurring feature of international criminal proceedings. Some of it will be justified. Some of it will be opportunistic. The law's task is not to eliminate doubt, but to discipline it.

The Rome Statute already contains the materials for a principled response. Article 69(4) provides the primary reliability and admissibility framework. Article 64 supplies case-management authority. Article 67 protects the accused against unfair burdens and preserves equality of arms. Article 66 keeps the prosecution's ultimate burden intact. Rules 63, 64 and 79 show that ICC procedure can combine flexibility, fair-trial protection and procedural organisation.

The article has proposed a three-stage framework. The proponent of digital evidence must provide a proportionate baseline showing of authenticity, provenance and reliability. The objector must identify case-specific reasons for synthetic doubt before enhanced review is triggered. The Chamber must then conduct proportionate reliability review, selecting the least intrusive response consistent with fairness, probative value and the integrity of proceedings.

AI-related provenance information and synthetic-media detector outputs supplement conventional authentication: provenance credentials may strengthen the foundation and detector alerts may trigger scrutiny, but neither is dispositive.

This approach rejects both extremes. A party should not be able to neutralise a well-corroborated video merely by saying that deepfakes exist. Nor should a Chamber rely uncritically on a central digital exhibit when concrete indicia of manipulation, provenance failure or forensic uncertainty have been identified. Synthetic doubt should be treated as a structured procedural trigger, not as an evidentiary veto and not as a technological anxiety to be ignored. That is the best way to preserve the integrity of digital proof before the International Criminal Court.

**Funding:**
No funding was received for this work.

**Conflict of interest:**
The author declares no conflict of interest.